# Price discrimination, algorithmic decision-making, and European non-discrimination law

## Prof. Dr. Frederik Zuiderveen Borgesius[1]
### iHub, Radboud University, The Netherlands

**Abstract -** Our society can benefit immensely from algorithmic decision-making and similar types of artificial intelligence. But algorithmic decision-making can also have discriminatory effects. This paper examines that problem, using online price differentiation as an example of algorithmic decision-making. With online price differentiation, a company charges different people different prices for identical products, based on information the company has about those people. The main question in this paper is: to what extent can non-discrimination law protect people against online price differentiation? The paper shows that online price differentiation and algorithmic decision-making could lead to indirect discrimination, for instance harming people with a certain ethnicity. Indirect discrimination occurs when a practice is neutral at first glance, but ends up discriminating against people with a protected characteristic, such as ethnicity. In principle, non-discrimination law prohibits indirect discrimination. The paper also shows, however, that non-discrimination law has flaws when applied to algorithmic decision-making. For instance, algorithmic discrimination can remain hidden: people may not realise that they are being discriminated against. And many types of unfair – some might say discriminatory – algorithmic decisions are outside the scope of current non-discrimination law.

[1] Professor of Law and ICT, iHub and Institute for Computing and Information Sciences (iCIS), Radboud University Nijmegen (the Netherlands). Contact: Frederikzb[at]cs.ru.nl. I would like to thank Noël Bangma, Janneke Gerards, Kees Groenendijk, Chantal Mak, Karin de Vries and the journal's anonymous peer reviewers for their valuable suggestions. Any errors are my own.

# 1 Introduction

This paper examines the interplay between, on the one hand, non-discrimination law, and on the other hand, algorithmic decision-making and similar types of artificial intelligence. Some organisations use thousands of data points to take algorithmic decisions: lenders can use algorithms to set, automatically, interest rates for individual consumers, or refuse to lend to them. Insurers can adjust premiums to individual consumers, or deny them insurance. Online stores can personalise offers, or sell the same good to different consumers for different prices. Such algorithmic decision-making can improve services, enhance efficiency, and foster economic growth. However, algorithmic decision-making may also threaten human rights, such as the right to non-discrimination.

Taking online price differentiation (also called price discrimination) as an example, this paper explores how non-discrimination law applies to algorithmic decision-making. With online price differentiation, an online store charges, automatically, different prices to different consumers for the same product, based on information about individual consumers. Surveys show that the general public dislikes most forms of online price differentiation. [2] Some consumers find online price differentiation 'very discriminatory'.[3] But is it, from a legal perspective?

The paper's main question is: to what extent can non-discrimination law protect people against online price differentiation? The paper focuses on EU non-discrimination law, rather than on national law. Discrimination on the basis of nationality is outside the scope of this paper.[4] The paper only discusses non-discrimination law; outside scope are, for instance, consumer law,[5] competition law,[6] and data protection law.[7] The paper speaks of 'discrimination' when referring to objectionable or illegal discrimination, for instance on the basis of skin colour. The paper uses the word 'differentiation' for discrimination, or making distinctions, in a neutral sense.

While the relevance of EU non-discrimination law for online price differentiation and algorithmic decision-making has been discussed by some authors,[8] this paper provides a more detailed analysis, and focuses on online price differentiation.[9]

An introduction to algorithmic decision-making and online price differentiation is given in section 2. Next, section 3 introduces non-discrimination law and the concepts

---

[3] Jennifer Valentino-Devries, Jeremy Singer-Vine, Ashkan Soltani, 'Websites Vary Prices, Deals Based on Users' Information', Wall Street Journal, 23 December 2012. http://online.wsj.com/article/SB10001424127887323777204578189391813881534.html

[4] Outside scope are, for instance, the Services Directive (2006/123/EC) and the Geo Blocking Regulation (2018/302). The Racial Equality Directive 2000/43/EC states that it does not cover difference of treatment based on nationality (article 3(2)).

[5] See on price differentiation and consumer protection law: Tycho de Graaf, *Consequences of Nullifying an Agreement on Account of Personalised Pricing*, Journal of European Consumer and Market Law 8.5 (2019), 184-193.

[6] See on price differentiation and competition law: Inge Graef, *Algorithms and Fairness: What Role for Competition Law in Targeting Price Discrimination Towards End Consumers?*, Columbia Journal of European Law. 24, 3, p. 541-559.

[7] See on price differentiation and data protection law: Richard Steppe, *Online price discrimination and personal data: A General Data Protection Regulation perspective* (2017) 33(6) Computer Law and Security Review 768; Frederik Zuiderveen Borgesius and Joost Poort, *Online Price Differentiation and EU Data Privacy Law*, Journal of Consumer Policy, 2017, 40(3), 347-366.

[8] See for instance: Philipp Hacker, *Teaching fairness to artificial intelligence: Existing and novel strategies against algorithmic discrimination under EU law*, (2018) 55 Common Market Law Review, Issue 4, pp. 1143–1185; Sandra Wachter, *Affinity Profiling and Discrimination by Association in Online Behavioural Advertising*, version 28 Sep 2019, https://ssrn.com/abstract=3388639. In the US, see Solon Barocas and Andrew Selbst, *Big Data's disparate impact* (2016) 104 Calif Law Rev 671.

[9] Drechsler and Benito Sánchez apply non-discrimination law to online price differentiation; my paper gives a more detailed analysis. Laura Drechsler and Juan Carlos Benito Sánchez, *The Price Is (Not) Right: Data Protection and Discrimination in the Age of Pricing Algorithms* (2018) 9(3) European Journal of Law and Technology

'direct' and 'indirect' discrimination. Section 4 applies non-discrimination law to online price differentiation.

I show that, in some circumstances, online price differentiation leads to indirect discrimination, which is prohibited. However, the concept of indirect discrimination is often difficult to apply in practice. A subtle proportionality test is required to assess whether a practice constitutes prohibited indirect discrimination. Moreover, many types of algorithmic decisions can be seen as unfair, some might say discriminatory, while they remain outside the scope of non-discrimination law. Section 5 and 6 give suggestions for further research and offer concluding thoughts.

## 2    Algorithmic decision-making and price differentiation

What is algorithmic decision-making? An algorithm can be described as a 'precisely specified series of instructions for performing some concrete task'.[10] For the purposes of this paper, 'decision' simply means the output of that algorithm. Loosely speaking, one could think of an algorithm as a computer program. Algorithmic decision-making is indispensable in a modern society. Algorithms provide us routes on the roads, search results on the web, etc.

Algorithms can also be used to make decisions about individuals or groups. For example, algorithms can help schools by predicting which students may need extra help, and help the police by predicting where crime may occur, or who may commit crime.[11]

But with algorithmic decision-making, there is a risk or unfair and illegal discrimination, for instance when algorithms are trained on data reflecting biased human decisions.[12] Many scholars and policymakers worry about discriminatory effects of algorithmic decision-making.[13] To keep the discussion manageable, we focus on one type of algorithmic decision-making in this paper: online price differentiation.

---

[10] Michael Kearns and Aaron Roth, The ethical algorithm (Oxford: Oxford University Press), 2019, p. 4. See also Paul Dourish, *Algorithms and their others: Algorithmic culture in context* (2016) 3(2) Big Data and Society.

[11] See for an overview of algorithmic decision-making in Europe: AlgorithmWatch, *Automating Society: Taking Stock of Automated Decision-Making in the EU – A report by AlgorithmWatch in cooperation with Bertelsmann Stiftung, supported by the Open Society Foundations*, Berlin 2019 www.algorithmwatch.org/automating-society

With online price differentiation., an online store differentiates the price for identical products based on information that a store has about a consumer. Such practices could also be called 'personalised pricing', 'algorithmic pricing', or 'price discrimination'. This paper mainly uses the phrase 'price differentiation'.

Many people have probably experienced seeing online advertising for products that they looked at earlier on the web. You need a lawnmower, you visit a few websites about lawnmowers, and later you see advertising for lawnmowers on other sites. For such behavioural advertising, companies monitor people's online behaviour to use the collected information to show people individually targeted advertisements. Similar technology could be used to present different prices to different internet users.

Economic literature shows that price differentiation typically leads to more profit for companies.[14] And marketing literature has said for at least two decades that stores should use online price differentiation. As one book puts it: 'Since some customers are willing to pay more than others for a product, companies that set only one price end up in pricing catch-22. Profits are foregone from customers who would have paid more, while additional customers would have purchased if the price were lower'.[15]

---

[12] See Solon Barocas and Andrew Selbst, *Big Data's disparate impact* (2016) 104 Calif Law Rev 671.; Frederik Zuiderveen Borgesius, *Discrimination, artificial intelligence, and algorithmic decision-making*, report for the European Commission against Racism and Intolerance (ECRI), Council of Europe, 2019, https://www.coe.int/en/web/artificial-intelligence/-/news-of-the-european-commission-against-racism-and-intolerance-ecri-

[13] See for instance the ACM Conference on Fairness, Accountability, and Transparency (ACM FAccT) https://facctconference.org.

[14] See for a summary of the economic literature, with further references: Frederik Zuiderveen Borgesius and Joost Poort, *Online Price Differentiation and EU Data Privacy Law*, Journal of Consumer Policy, 2017, 40(3), 347-366.

[15] Ravi Mohammed, *The 1% windfall: how successful companies use price to profit and grow,* p. 109 (New York: Harper Collins 2010). In that sentence, Mohammed is referring to price differentiation in general; not to personalised pricing.

An online store can see where consumers (website visitors) are located on the basis of their IP addresses.[16] Research from 2012 showed that several online stores in the US charge people from different areas different prices.[17] One online store charged more to people in the countryside than to people who live in a large city. In the countryside, somebody who does not like the online price might have to travel for hours to visit a competing (brick and mortar) store. Therefore, the online store charged relatively high prices to customers in the countryside. In large cities, people can easily go to a competitor. Therefore, the store offered cheaper prices in cities, to ensure that consumers choose their store over a competitor. On average, people in the countryside are poorer than people in large cities in the US. Hence, the pricing scheme had the effect, probably unintentionally, that poorer people paid, on average, higher prices.[18] In Europe too, some online stores charge people from different regions different prices.[19]

Such price differentiation on the basis of a consumer's location can lead to higher prices for people with a certain skin colour. For instance, in the U.S., the tutoring service Princeton Review charged different prices to people in different neighbourhoods. This pricing scheme resulted in higher prices for neighbourhoods where many people of Asian descent live.[20] Presumably, the company did not aim to discriminate on the basis of skin colour. Perhaps the company tested different prices for different areas. After such tests, a company can charge higher prices in areas where consumers continue to buy, even at higher prices. Hence, the company probably wanted to improve profits, which had the side effect of harming people from an Asian background. (We will see in section 4.2 that for the law it is not always relevant whether a company discriminates on purpose or by accident.)

---

[16] The IP address does not provide the user's location in 100% of the cases. For instance, people can change their IP address using a VPN-service.
[17] Jennifer Valentino-Devries, Jeremy Singer-Vine, Ashkan Soltani, 'Websites Vary Prices, Deals Based on Users' Information', Wall Street Journal, 23 December 2012. http://online.wsj.com/article/SB10001424127887323777204578189391813881534.html

[18] Jennifer Valentino-Devries, Jeremy Singer-Vine, Ashkan Soltani, 'Websites Vary Prices, Deals Based on Users' Information', Wall Street Journal, 23 December 2012. http://online.wsj.com/article/SB10001424127887323777204578189391813881534.html
[19] Jakub Mikians et al., *Crowd-assisted search for price differentiation in e-commerce: first results*, Proceedings of the ninth ACM conference on Emerging networking experiments and technologies (p. 1–6), 2013.
[20] Jeff Larson, Surya Mattu, Julia Angwin, *Unintended Consequences of Geographic Targeting*, Technology Science. 2015090103. September 01, 2015. https://techscience.org/a/2015090103

Online stores can also adapt prices to other consumer characteristics than the consumer's location. For instance, some stores charge different prices to cell phone users than to computer users.[21] Stores could also charge different prices to Apple users than to PC users.[22] In sum, modern technology enables companies to use narrowly targeted price differentiation. New developments such as big data, machine learning, and artificial intelligence give companies even more possibilities for price differentiation.[23]

This paper focuses on one type of online price differentiation, where prices are adapted to information about individual consumers (personalised pricing). Outside the scope of this paper are, for instance, discounts when buying higher quantities, or higher ticket prices when an airplane is almost sold out. Risk-based pricing is also outside the scope of this paper. An example of risk-based pricing is an insurance company that charges more to certain groups of drivers, because the company expects that those drivers will have more accidents.

While there is evidence of online price differentiation in the wild, it appears that most stores do not adapt prices to consumers (yet).[24] Several explanations are possible. First, some stores may refrain from price differentiation because they fear the reaction of consumers. Second, some stores may still lack the skills or data to adapt prices to consumers with sufficient precision. However, that might change. Third, perhaps price differentiation happens – or will happen – without being noticed. Stores could differentiate prices in ways that are hard to detect. For example, a store could advertise one price on its website but email each consumer a different discount coupon. Many commentators expect that price differentiation will become more common, even in brick and mortar stores.[25] Now we turn to law; in particular to non-discrimination law.

---

[21] Aniko Hannak et al., *Measuring price discrimination and steering on e-commerce web sites*, Proceedings of the 2014 Conference on Internet Measurement Conference, 305-318.

[22] There is no evidence of Apple users paying more for the same goods than PC users, but the technology is there. One site showed more expensive hotels to Apple users than to PC users. Dana Mattioli, 'On Orbitz, Mac Users Steered to Pricier Hotels', *Wall Street Journal,* 23 August 2012.

[23] Executive Office of the President of the United States (Council of Economic Advisors) 'Big Data and Differential Pricing', 2015.

[24] For instance, there is no hard evidence of airlines using personalised pricing. See Thomas Vissers et al., *Crying wolf? On the price discrimination of online airline tickets*, 7th Workshop on Hot Topics in Privacy Enhancing Technologies (HotPETs 2014).

## 3 Non-discrimination law

### 3.1 EU discrimination directives

The right to non-discrimination is included in many human rights treaties such as the European Convention on Human Rights (1950),[26] the International Convention on the Elimination of all Forms of Racial Discrimination (1965),[27] the International Covenant on Civil and Political Rights (1966),[28] and the Charter of Fundamental Rights of the European Union (2000).[29]

The EU has also adopted rules combatting discrimination.[30] The four most relevant discrimination directives are the following. (i) The Racial Equality Directive (2000) prohibits discrimination on the basis or racial or ethnic origin in many contexts.[31] (ii) The Employment Equality Directive (2000) prohibits discrimination on the grounds of religion or belief, disability, age, or sexual orientation in the employment context.[32] (iii) The Gender Goods and Services Directive (2004) prohibits discrimination on the basis of gender in the context of the supply of goods and services.[33] The Recast Gender

---

[25] See Andrew Odlyzko, *Privacy, economics, and price discrimination on the Internet,* Proceedings of the 5th international conference on Electronic commerce (ACM) 355, 2003.
[26] Article 14 of the European Convention on Human Rights.
[27] See in particular article 1-7 of the International Convention on the Elimination of all Forms of Racial Discrimination.
[28] Article 2 and 26 of the International Covenant on Civil and Political Rights.
[29] Article 21 of the Charter of Fundamental Rights of the European Union. See also article 19 of the Treaty on the Functioning of the European Union.
[30] See for the history of EU discrimination law: Mark Bell, *Anti-discrimination law and the European Union,* chapter 2, p. 32-53 (Oxford: Oxford University Press 2002).
[31] Council Directive 2000/43/EC Implementing the Principle of Equal Treatment Between Persons Irrespective of Racial or Ethnic Origin, 2000 OJ L 180/22.
[32] Council Directive 2000/78/EC Establishing a General Framework tor Equal Treatment in Employment and Occupation, 2000 OJ L 303/16.
[33] Council Directive 2004/113/EC Implementing the principle of Equal Treatment between Men and Women in the Access to and Supply of Goods and Services, 2004 OJ L 373/37.

Equality Directive (2006) prohibits discrimination on the basis of gender in the employment context.[34]

Member states must implement a directive in their national law,[35] and they can adopt stricter rules than required by the discrimination directives. This paper focuses on the directives, rather than on national implementation laws.

## 3.2 Direct and indirect discrimination

EU law prohibits both 'direct' and 'indirect' discrimination. Both concepts are defined almost the same in the four discrimination directives. Direct discrimination is defined as follows in the Racial Equality Directive:

> Direct discrimination shall be taken to occur where one person is treated less favourably than another is, has been or would be treated in a comparable situation on grounds of ethnic or ethnic origin.[36]

Roughly summarised, direct discrimination means that organisations discriminate against people on the basis of a protected characteristic, such as ethnic origin. Direct discrimination occurs, for example, when a company says it will not recruit employees with a certain skin colour.[37]

'Indirect' discrimination is the second category of prohibited discrimination.[38] We make a brief detour to US law, because the US Supreme Court developed the concept of indirect discrimination, called 'disparate impact' in the US (in the case Griggs v Duke Power, 1971). The Civil Rights Act of 1964 prohibited intentional (direct) discrimination based on race, colour, religion, sex, or national origin. Before that Act entered into force, an employer, Duke Power, openly discriminated against African Americans.[39] After the Act entered into force, the employer required a high school diploma and a certain IQ test score from applicants for certain jobs. These IQ requirements, however, did not directly relate to what was needed for the jobs. The Court ruled that the employer's requirements were discriminatory because African Americans scored lower on IQ tests and less often possessed a high school diploma, because of 'inferior education in segregated schools'.[40]

---

[34] Directive 2006/54/EC of the European Parliament and of the Council on the implementation of the Principle of Equal Opportunities and Equal Treatment of Men and Women in Matters of Employment and Occupation (Recast), 2006 OJ L 204/23.
[35] A directive is binding for the EU member states regarding the result to be achieved, but leaves the choice of form and methods to the member states. See article 288 of the Treaty on The Functioning of the European Union, consolidated version https://eur-lex.europa.eu/resource.html?uri=cellar:2bf140bf-a3f8-4ab2-b506-fd71826e6da6.0023.02/DOC_2&format=PDF.
[36] Article 2(2)(a) of the Racial Equality Directive 2000/43/EC; capitalisation and punctuation adapted.
[37] CJEU, C-54/07, 10 July 2008, Centrum voor gelijkheid van kansen en voor racismebestrijding v Firma Feryn NV.
[38] The EU directives also prohibit harassment and the instruction to discriminate. See e.g. article 2 of the Racial Equality Directive 2000/43/EC.

The Supreme Court said: 'The Act proscribes not only overt discrimination but also practices that are *fair in form, but discriminatory in operation*. The touchstone is business necessity. If an employment practice, which operates to exclude Negroes cannot be shown to be related to job performance, the practice is prohibited.'[41] The Court added that discriminatory intent is not relevant for such disparate impact.[42] US law on indirect discrimination influenced EU law.[43]

Roughly speaking, indirect discrimination occurs when a practice is neutral at first glance but ends up discriminating against people with a certain ethnic origin (or another protected characteristic).[44] Hence, indirect discrimination focuses on the effects of a practice; not on the intention of the alleged discriminator. The Racial Equality Directive defines indirect discrimination as follows:

---

[39] David Garrow, *Toward a Definitive History of Griggs v. Duke Power Co*, (2014) 67 Vand.L.Rev. 197.
[40] US Supreme Court, Griggs v. Duke Power Co., 401 U.S. 424 (1971), p. 430.
[41] US Supreme Court, Griggs v. Duke Power Co., 401 U.S. 424 (1971), p. 431, emphasis added.
[42] US Supreme Court, Griggs v. Duke Power Co., 401 U.S. 424 (1971), p. 432. The prohibition of disparate impact was codified in the Civil Rights Act of 1991.
[43] Sandra Fredman, *Discrimination law*, p. 178, (Oxford: Oxford University Press 2011); Christa Tobler, *Limits and potential of the concept of indirect discrimination* (report for the Directorate-General for Employment, Social Affairs and Inclusion, European Commission), 2008 https://publications.europa.eu/en/publication-detail/-/publication/aa081c13-197b-41c5-a93a-a1638e886e61/language-en, p. 23-24.
[44] See generally on the concept of indirect discrimination: Christa Tobler, *Indirect discrimination: a case study into the development of the legal concept of indirect discrimination under EC law*, vol 10 (Antwerp: Intersentia 2005); Evelyn Ellis and Philippa Watson, p. 148-155, *EU anti-discrimination law* (Oxford: Oxford University Press 2012).

> Indirect discrimination shall be taken to occur where [i] an apparently neutral provision, criterion or practice [ii] would put persons of a racial or ethnic origin at a particular disadvantage compared with other persons, [iii] unless that provision, criterion or practice is objectively justified by a legitimate aim and the means of achieving that aim are appropriate and necessary.[45]

The distinction between direct and indirect discrimination has important legal implications. Direct discrimination cannot be justified – apart from some explicitly defined narrow exceptions.[46] For instance, a women's clothing brand can recruit only women as models for its advertising, without breaching non-discrimination law.[47]

Indirect discrimination can be justified by invoking an objective justification. The possibility for justification is part of the definition of indirect discrimination: 'unless that provision, criterion or practice is objectively justified'.[48] Later in this paper (section 4.3), we discuss the possibility of justifying indirect discrimination in more detail, when we apply non-discrimination law to online price differentiation.

## 4 Applying non-discrimination law to price differentiation

### 4.1 Direct discrimination

Below we apply the general rules from the Racial Equality Directive to online price differentiation. Suppose that an online book store adapts the prices of its books to the consumer's location (based on his or her IP address). The store differentiates its prices to improve profit. It turns out that people pay, on average, 20% extra if they live in streets where a majority of the people have a Roma background. We assume that the store does not intend to discriminate against Roma, and that the prices are independent of postage costs, taxes, etc.

---

[45] Article 2(2)(b) of the Racial Equality Directive 2000/43/EC; capitalisation and punctuation adapted; numbering added.
[46] Direct discrimination can be justified in 'very limited circumstances' (recital 18), in particular in the case of genuine occupational requirements (article 4 of the Racial Equality Directive 2000/43/EC).
[47] See article 14(2) of the Gender Goods and Services Directive 2006/54/EC.
[48] Article 2(2)(b) of the Racial Equality Directive 2000/43/EC. See about 'objective justification': Hugh Collins, *Justice for Foxes: Fundamental Rights and Justification of Indirect Discrimination* in Hugh Collins and Tarunabh Khaitan (eds), *Foundations of Indirect Discrimination Law* (Oxford: Hart Publishing 2018).

Does this price differentiation fall within the scope of the Racial Equality Directive? The directive applies in 'both the public and private sectors', and to 'access to and supply of goods and services which are available to the public'.[49] The store sells goods to the public; the Racial Equality Directive thus applies.[50]

The Racial Equality Directive prohibits discrimination based on 'racial or ethnic origin.'[51] The directive does not define the words 'racial' and 'ethnic' origin.[52] The Directive's preamble notes that the EU 'rejects theories which attempt to determine the existence of separate human races'.[53] The Court of Justice of the European Union (CJEU) says that the phrase 'racial or ethnic origin' must be interpreted widely.[54] And the CJEU has confirmed that the Racial Equality Directive protects Roma against discrimination.[55]

There are not many judgments by the CJEU on the Racial Equality Directive, but there is a Grand Chamber judgment from 2015 regarding a company that adapted its practices (but not prices) to the customers' location: *Chez/Nikolova*.[56] Roughly summarised, an electricity company, called Chez Razpredelenie Bulgaria AD, installed electricity meters at a height of 6 meters in districts densely populated by Roma. In other districts, the company placed electricity meters at a height of 1.70 meters. The company said that it installed the meters extra high to prevent tampering with the meters and electricity theft.

---

[49] Article 3(1)(h) of the Racial Equality Directive 2000/43/EC.

[50] Article 3(1)(h) of the Racial Equality Directive 2000/43/EC. The CJEU adds that the scope of the directive must not be interpreted restrictively. CJEU (Grand chamber), Case C-83/14, 16 July 2015, Chez/Nikolova, ECLI:EU:C:2015:480, par. 42.

[51] Article 2(1) of the Racial Equality Directive 2000/43/EC. The Directive concerns minimum harmonisation; Member States can provide more protection than required by the directive. See CJEU (Grand chamber), Case C-83/14, 16 July 2015, Chez/Nikolova, ECLI:EU:C:2015:480, par. 67.

[52] See generally: Lilla Farkas, *The meaning of racial or ethnic origin in EU law: between stereotypes and identities, European network of legal experts in gender equality and non-discrimination*, report for European Commission, Directorate-General for Justice and Consumers (January 2017) https://www.equalitylaw.eu/downloads/4030-the-meaning-of-racial-or-ethinic-origin-in-eu-law-between-stereotypes-and-identities   See on the word 'racial' also recital 16 of the Racial Equality Directive.

[53] Recital 6 of the Racial Equality Directive 2000/43/EC.

[54] CJEU (Grand chamber), Case C-83/14, 16 July 2015, Chez/Nikolova, ECLI:EU:C:2015:480, par. 56. But somebody's country of birth does not, by itself, disclose that person's ethnic origin, said the Court in a later judgment: CJEU, Case C-688/15, 6 April 2017, Jyske Finans, ECLI:EU:C:2018:209, par. 20. See for a critical discussion of the CJEU's reasoning in that case: Shreya Atrey, *Race discrimination in EU law after Jyske Finans* (2018) 55(2) Common Market Law Review 625.

[55] CJEU (Grand chamber), Case C-83/14, 16 July 2015, Chez/Nikolova, ECLI:EU:C:2015:480.

Ms. Nikolova lives in a predominantly Roma district. She complained to the Bulgarian Commission for Protection against Discrimination that the meter was installed too high for her to be able to check her electricity usage. After some legal wrangling, a Bulgarian judge asked the CJEU for advice on how to interpret the Racial Equality Directive.

Ms. Nikolova declared that she does not have a Roma background. The CJEU said that her own background was not relevant in this case; the prohibition of discrimination in the Racial Equality Directive would protect her regardless of whether she belonged to an ethnic minority.[57]

We return to our price differentiation example. Does the price differentiation entail prohibited discrimination? First, we discuss direct discrimination. The Racial Equality Directive says that direct discrimination occurs 'where one person is treated less favourably than another is, has been or would be treated in a comparable situation on grounds of racial or ethnic origin.'[58] The CJEU said in *Chez/Nikolova*: 'a measure such as the practice at issue constitutes direct discrimination (…) if that measure proves to have been introduced and/or maintained for reasons relating to the ethnic origin common to most of the inhabitants of the district concerned'.[59] (Whether there was 'direct' discrimination was for the national judge to decide.)

---

[56] CJEU (Grand chamber), Case C-83/14, 16 July 2015, Chez/Nikolova, ECLI:EU:C:2015:480.
[57] CJEU (Grand chamber), Case C-83/14, 16 July 2015, Chez/Nikolova, ECLI:EU:C:2015:480, par. 91. See also par. 95.
[58] Article 2(2)(a) of the Racial Equality Directive 2000/43/EC; capitalisation and punctuation adapted.
[59] CJEU (Grand chamber), Case C-83/14, 16 July 2015, Chez/Nikolova, ECLI:EU:C:2015:480, par. 91. See also par. 95.

However, in our example, the store differentiates prices for profit – not for reasons related to ethnic origin. Hence, our example does not concern direct discrimination.[60] Apart from that, there don't seem to be real-life examples of automated price discrimination that entails direct discrimination. We turn to indirect discrimination in the next section.

## 4.2    Indirect discrimination

Does our price differentiation example entail 'indirect' discrimination? As already seen, indirect discrimination happens when an 'apparently neutral' practice puts people of a certain ethnic origin at a disadvantage compared with other persons, unless the company can justify the practice.

(i) The first question is: is price differentiation on the basis of location an 'apparently neutral' practice?[61] An 'apparently neutral' practice, says the CJEU, means a 'practice which is worded or applied, ostensibly, in a neutral manner, that is to say, having regard to factors different from and not equivalent to the protected characteristic'.[62] The price differentiation on the basis of a consumer's location is such an apparently neutral practice.[63]

(ii) The next question is: would the price differentiation 'put persons of a racial or ethnic origin at a particular disadvantage compared with other persons'?[64] The word 'disadvantage' must be interpreted widely, says the CJEU:[65] 'the concept of "particular disadvantage" (…) does not refer to serious, obvious or particularly significant cases of inequality'.[66] The CJEU adds that 'no particular degree of seriousness is required' for the prohibition of indirect discrimination to apply.[67]

---

[60] There would be a case of direct discrimination if a store would put up a sign that says, for instance, 'Roma pay 10% extra'.
[61] In other words, is the practice neutral 'at first glance'? CJEU (Grand chamber), Case C-83/14, 16 July 2015, Chez/Nikolova, ECLI:EU:C:2015:480, par 93.
[62] CJEU (Grand chamber), Case C-83/14, 16 July 2015, Chez/Nikolova, ECLI:EU:C:2015:480, par. 109. See also par. 94.
[63] In *Chez/Nikolova,* the CJEU came to a similar conclusion about a company that adapted its practices to different districts. CJEU (Grand chamber), Case C-83/14, 16 July 2015, Chez/Nikolova, ECLI:EU:C:2015:480, par. 109. See also par. 91 and par. 106.
[64] Article 2(2)(b) of the Racial Equality Directive 2000/43/EC.
[65] CJEU (Grand chamber), Case C-83/14, 16 July 2015, Chez/Nikolova, ECLI:EU:C:2015:480, par. 67; par. 69. See also par. 99 and 103.

In our example, Roma pay higher prices compared with others (who live in a different neighbourhood). Paying extra is a 'disadvantage'. Hence: in our example, the 'particular disadvantage' criterion is met.

However, consumers may not realise that they see different prices than other people. The lack of transparency of price differentiation (and of algorithmic decision-making in general) makes it harder to detect discrimination.[68] Additionally, even if, for instance, a Roma consumer discovered that he or she paid a different price than others, that consumer would still not know that Roma pay more in general. Consumers who do not realise that they are discriminated against will not exercise their right to non-discrimination.

But suppose that a consumer suspects that Roma pay more, and suppose that the consumer goes to court. EU non-discrimination law reverses the burden of proof. When a consumer brings a case on discrimination to court and provides 'facts from which it may be presumed that there has been direct or indirect discrimination, it shall be for the respondent to prove that there has been no breach of the principle of equal treatment.'[69]

However, the consumer must still show that discrimination is likely. As Farkas and O'Farrel explain, 'The reversal of the burden of proof does not mean that plaintiffs are exempt from convincing the court that they have a case: a set of facts that call for an explanation. In order to reverse the burden of proof they must first establish a prima facie case, in other words convince the court of the likeliness or probability that they suffered discrimination.'[70] Consumers can use statistics to show that discrimination is likely.[71] However, it would be difficult for consumers to obtain statistics to prove such an (indirectly) discriminatory effect.[72]

---

[66] CJEU (Grand chamber), Case C-83/14, 16 July 2015, Chez/Nikolova, ECLI:EU:C:2015:480, par. 109. See also par. 67, 69, and 99.
[67] CJEU (Grand chamber), Case C-83/14, 16 July 2015, Chez/Nikolova, ECLI:EU:C:2015:480, par 103.
[68] Perhaps consumers could use their data protection rights to obtain more information about how prices are set. See Laura Drechsler and Juan Carlos Benito Sánchez, *The Price Is (Not) Right: Data Protection and Discrimination in the Age of Pricing Algorithms* (2018) 9(3) European Journal of Law and Technology. See also Sandra Wachter, *Affinity Profiling and Discrimination by Association in Online Behavioural Advertising*, version 28 Sep 2019, https://ssrn.com/abstract=3388639, p. 42; p. 69.
[69] Article 8(1) of the Racial Equality Directive.

A store cannot justify indirect discrimination by proving that it did not mean to discriminate on the basis of ethnic origin.[73] For indirect discrimination, it is not relevant whether the store intended to discriminate against Roma. Accidental indirect discrimination is still prohibited.

To sum up: online price differentiation can involve two elements of indirect discrimination: (i) an apparently neutral practice, (ii) that puts people of a certain ethnic origin at a disadvantage. However, it might be difficult to discover the price differentiation. The next element of indirect discrimination is (iii) a lack of objective justification; we turn to that element now.

## 4.3    Objective justification

Is the price differentiation 'objectively justified by a legitimate aim'? And are 'the means of achieving that aim (…) appropriate and necessary'?[74] Such questions are often difficult to answer; the answer depends on all circumstances of a case.[75] The CJEU says that 'the concept of objective justification must be interpreted strictly.'[76]

---

The burden of proof lies on the defendant: the store would have to prove that it has an objective justification.[77]

First: does the price differentiation serve a 'legitimate aim'? This phrase is not defined in the Racial Equality Directive.[78] The CJEU tends to leave it to national judges to assess whether there is a legitimate aim. The case law of the CJEU gives thus limited guidance on which aims should be considered legitimate.[79] Therefore, the analysis below remains somewhat tentative.

In our example, the online store adapts prices to the consumer's location, to improve profit. Several arguments suggest that profit improvement is a legitimate aim. First, profit improvement is, in principle, lawful. Indeed, profit making lies at the core of our market economies. Second, the store can invoke a fundamental right, namely its 'freedom to conduct a business in accordance with Community law and national laws and practices' (article 16 of the Charter of Fundamental Rights of the European Union).

The 2017 *Achbita* case illustrates the reasoning by the CJEU about a 'legitimate aim' in the context of indirect discrimination. Roughly summarised, the case concerned a company that offered reception services for clients in the public and the private sector. The company prohibited, in an internal rule, its employees to wear visible political, philosophical, or religious signs. The company fired Ms. Achbite for wearing an Islamic headscarf. She complained about discrimination. A Belgian judge asked advice to the CJEU about how to interpret the concepts 'direct' and 'indirect' discrimination.

The CJEU said that the company's prohibition on wearing an Islamic headscarf did not constitute *direct* discrimination based on religion or belief.[80] But the company's internal rule could constitute *indirect* discrimination if it 'results, in fact, in persons adhering to a particular religion or belief being put at a particular disadvantage'.[81]

---

[77] Article 8(1) of the Racial Equality Directive 2000/43/EC. See also Christa Tobler, *Limits and potential of the concept of indirect discrimination* (report for the Directorate-General for Employment, Social Affairs and Inclusion, European Commission), 2008 https://publications.europa.eu/en/publication-detail/-/publication/aa081c13-197b-41c5-a93a-a1638e886e61/language-en , p. 32.

[78] Idem, p. 32.

[79] Idem, p. 33.

[80] CJEU, Case C-157/15 Achbita and Centrum voor Gelijkheid van Kansen en voor Racismebestrijding v. G4 S Secure Solutions NV, EU:C:2017:203, dictum. The case concerned the Employment Equality Directive (2000/78/EC). The definitions of 'direct' and 'indirect' discrimination resemble those in the Racial Equality Directive.

The CJEU added that there could be an exception: if the company's internal rule 'is objectively justified by a legitimate aim, such as the pursuit by the employer, in its relations with its customers, of a policy of political, philosophical and religious neutrality'.[82] The CJEU connected the company's aim for a neutral image to the freedom to conduct a business: '[a]n employer's wish to project an image of neutrality towards customers relates to the freedom to conduct a business that is recognised in Article 16 of the Charter and is, in principle, legitimate (…)'.[83] Scholars have criticised the *Achbita* judgment for being too business-friendly.[84]

In her opinion for the *Achbita* case, Advocate General Kokott also emphasised the importance of the freedom to conduct a business:

> In a Union which regards itself as being committed to a social market economy (second sentence of Article 3(3) TEU [Treaty on European Union]) and seeks to achieve this in accordance with the requirements of an open market economy with free competition (Articles 119(1) TFEU and 120 TFEU [Treaty on the Functioning of the European Union]), the importance that

---

[81] CJEU, Case C-157/15 Achbita and Centrum voor Gelijkheid van Kansen en voor Racismebestrijding v. G4 S Secure Solutions NV, EU:C:2017:203, dictum.
[82] CJEU, Case C-157/15 Achbita and Centrum voor Gelijkheid van Kansen en voor Racismebestrijding v. G4 S Secure Solutions NV, EU:C:2017:203, dictum.
[83] CJEU, Case C-157/15 Achbita and Centrum voor Gelijkheid van Kansen en voor Racismebestrijding v. G4 S Secure Solutions NV, EU:C:2017:203, par. 38.
[84] See: Stéphanie Hennette-Vauchez, *Equality and the Market: the unhappy fate of religious discrimination in Europe: ECJ 14 March 2017, Case C-188/15, Asma Bougnaoui and ADDH v Micropole SA; ECJ 14 March 2017, Case C-157/15, Samira Achbita and Centrum voor gelijkheid van kansen en voor racismebestrijding v G4S Secure Solutions NV* (2017) 13(4) European Constitutional Law Review 744; Lucy Vickers, *Headscarves and the Court of Justice of the EU: Discrimination and Genuine Occupational Requirements* (2017) 3(3) International Labor Rights Case Law 413; Erica Howard, *Islamic headscarves and the CJEU: Achbita and Bougnaoui* (2017) 24(3) Maastricht Journal of European and Comparative Law 348; Mark Bell, *Leaving Religion at the Door? The European Court of Justice and Religious Symbols in the Workplace* (2017) 17(4) Human Rights Law Review 784.

> attaches to the freedom to conduct a business is not to be underestimated.[85]

Advocate General Kokott added that a company has the 'right, in principle, to determine (…) in what form its products and services are offered.'[86] Case law of the CJEU confirms that the freedom to conduct a business (article 16 of the Charter) 'covers, inter alia, freedom of contract'.[87] Freedom of contract implies that parties are, in principle, free to set prices.[88]

In sum, there are several arguments in favour of concluding that, under current EU law, a company can invoke profit improvement as a legitimate aim. More generally, scholars conclude from the case law of the CJEU that meeting the 'legitimate aim' requirement is usually rather easy.[89]

But there are also arguments against accepting profit improvement as a legitimate aim. First, there is no case law of the CJEU that explicitly states that improving profit is a legitimate aim. Second, profit improvement might be too vague to be characterised as a 'legitimate aim'. Examples of legitimate aims from case law are more specific, such as fraud prevention,[90] protecting people's health,[91] and promoting a neutral image.[92]

---

[85] AG Kokott, Opinion in Case C-157/15 Achbita and Centrum voor Gelijkheid van Kansen en voor Racismebestrijding v. G4 S Secure Solutions NV, ECLI:EU:C:2016:382, par. 134, internal citation omitted. See also the opinion of the AG in CJEU, C-188/15, 14 March 2017, Bougnaoui, par. 108 and 134.

[86] AG Kokott, Opinion in Case C-157/15 Achbita and Centrum voor Gelijkheid van Kansen en voor Racismebestrijding v. G4 S Secure Solutions NV, ECLI:EU:C:2016:382, par. 81, internal citation omitted.

[87] CJEU, C-426/11, Alemo-Herron and Others, EU:C:2013:521, par. 32.

[88] Jan Smits, *The Law of Contract,* in Jaap Hage, Antonia Waltermann, and Bram Akkermans (eds), *Introduction to Law* (Dordrecht: Springer 2017).

[89] Christa Tobler, *Limits and potential of the concept of indirect discrimination* (report for the Directorate-General for Employment, Social Affairs and Inclusion, European Commission), 2008 https://publications.europa.eu/en/publication-detail/-/publication/aa081c13-197b-41c5-a93a-a1638e886e61/language-en , p. 43.

[90] CJEU (Grand chamber), Case C-83/14, 16 July 2015, Chez/Nikolova, ECLI:EU:C:2015:480, par 113-114.

[91] CJEU (Grand chamber), Case C-83/14, 16 July 2015, Chez/Nikolova, ECLI:EU:C:2015:480, par 113-114.

[92] Case C-157/15 Achbita and Centrum voor Gelijkheid van Kansen en voor Racismebestrijding v. G4 S Secure Solutions NV, EU:C:2017:203.

Hence, the case law does not prove that a broad goal such as 'improving profit' can serve as a 'legitimate aim'.

There is a third argument against accepting profit improvement as a legitimate aim. The CJEU has said in discrimination cases that an EU member state cannot invoke budgetary considerations as a legitimate aim: 'although budgetary considerations may underlie a Member State's choice of social policy and influence the nature or scope of the social protection measures which it wishes to adopt, *they do not in themselves constitute an aim pursued by that policy and cannot therefore justify discrimination* (…).'[93] By analogy, it could be argued that profit improvement cannot be a legitimate aim for a company. Then again, some might respond that human rights requirements for states are higher than for companies.[94] Following that line of reasoning, a state cannot invoke 'budgetary considerations', but a company can invoke 'profit improvement'.

For argument's sake, let's assume that the store in our example passes the 'legitimate aim' hurdle. But having a legitimate aim is not enough to justify indirect discrimination; the means to achieve that aim must be proportionate and necessary. Hence, a proportionality test must be conducted.[95] The proportionality test has been described as follows by the CJEU:

> it is necessary to consider whether such measures (…) exceed the limits of what is [a] appropriate and [b] necessary in order to attain the objectives legitimately pursued by the [measures] in question; [c] when there is a choice between several appropriate measures, recourse must be had to the least

---

[93] CJEU, C-196/02, 10 March 2005, Nikoloudi, ECLI:EU:C:2005:141, par 53.
[94] See also Sandra Wachter, *Affinity Profiling and Discrimination by Association in Online Behavioural Advertising,* version 28 Sep 2019, https://ssrn.com/abstract=3388639, p. 50.
[95] See CJEU (Grand chamber), Case C-83/14, 16 July 2015, Chez/Nikolova, ECLI:EU:C:2015:480, par. 118. See also Advocate General Kokott, Case C-83/14, 12 March 2015, Chez/Nikolova, ECLI:EU:C:2015:170, par. 118-139.

> onerous, and the disadvantages caused must not be disproportionate to the aims pursued.[96]

In discrimination cases, the CJEU tends to leave the proportionality test to the national judge, because the test requires assessing all the facts and circumstances of a case.[97]

We apply the proportionality test to our price differentiation example. Is the price differentiation practice an (a) appropriate and (b) necessary means for improving profit?[98] We start with (a): does price differentiation constitute an appropriate, or suitable, means for the purpose of profit improvement? Literature in the fields of marketing and economics says that price differentiation helps to improve profit.[99] It can thus be assumed that price differentiation is an appropriate means.

(b) Is the price differentiation 'necessary' too?[100] In other words: to improve profit, could the company take other measures than price differentiation that would bring less harm to consumers with a Roma background? The answer depends on all the circumstances. For instance, it would be harder for a store to argue that differentiating prices is necessary, if competitors do not see the need to differentiate prices.[101]

In theory, the store could assess whether other pricing strategies would lead to less harm for Roma. And in theory, the store might be able to find a different pricing scheme which leads to the same profits, but causes less harm to Roma. However, in practice it could be difficult for the store to make such an assessment. For instance, the store might not have access to data about the ethnic origin of people in different areas.[102] Let's assume that the store passes the 'appropriate' and 'necessary' hurdles.

---

[96] CJEU, C-379/08 and C-380/08, ERG and Others, ECLI:EU:C:2010:127, par. 86. Numbering added by me. The ERG judgment does not concern discrimination, but I quote the judgment because it provides a concise description of the proportionality test.
[97] Christa Tobler, *Limits and potential of the concept of indirect discrimination* (report for the Directorate-General for Employment, Social Affairs and Inclusion, European Commission), 2008 https://publications.europa.eu/en/publication-detail/-/publication/aa081c13-197b-41c5-a93a-a1638e886e61/language-en , p. 35.
[98] Article 2(2)(b) of the Racial Equality Directive 2000/43/EC.
[99] See Frederik Zuiderveen Borgesius and Joost Poort, *Online Price Differentiation and EU Data Privacy Law*, Journal of Consumer Policy, 2017, 40(3), 347-366, with further references.
[100] See CJEU (Grand chamber), Case C-83/14, 16 July 2015, Chez/Nikolova, ECLI:EU:C:2015:480, par. 120.
[101] See CJEU (Grand chamber), Case C-83/14, 16 July 2015, Chez/Nikolova, ECLI:EU:C:2015:480, par. 121.

The next question (c) concerns proportionality *sensu strictu* (proportionality in the narrow sense). As the CJEU puts it: 'the referring court will also have to determine whether the disadvantages caused by the practice at issue are disproportionate to the aims pursued and whether that practice unduly prejudices the legitimate interests of the persons inhabiting the district concerned'.[103] In short, 'a fair balance must be struck between the conflicting interests'.[104]

Are the disadvantages caused by the price differentiation disproportionate to the aim (profit improvement), and does the price differentiation unduly prejudice the interests of Roma? As noted previously, the store's interests are related to its freedom to conduct a business, protected in the Charter of Fundamental Rights of the European Union.[105] But the freedom to conduct a business is not absolute. The CJEU explains that 'the freedom to conduct a business may be subject to a broad range of interventions on the part of public authorities that may limit the exercise of economic activity in the public interest'.[106]

Meanwhile, consumers have an interest in paying low prices for products. And surveys suggest that most consumers dislike online price differentiation.[107] Moreover, the EU

---

[102] The European Union has strict rules on the collection and use of personal data about racial and ethnic origin. See: Article 9 of the European Parliament and Council Regulation (EU) 2016/679 of 27 April 2016 on the protection of natural persons with regard to the processing of personal data and on the free movement of such data, and repealing Directive 95/46/EC (General Data Protection Regulation) [2016] OJ L 119/1.

[103] CJEU (Grand chamber), Case C-83/14, 16 July 2015, Chez/Nikolova, ECLI:EU:C:2015:480, par. 67; par. 123.

[104] AG Kokott, Opinion in Case C-157/15 Achbita and Centrum voor Gelijkheid van Kansen en voor Racismebestrijding v. G4 S Secure Solutions NV, ECLI:EU:C:2016:382, par. 112. However, sometimes the CJEU does not explicitly discuss proportionality *strictu sensu*. See on that point: Janneke Gerards, *Case note regarding Case C-157/15 Achbita and Centrum voor Gelijkheid van Kansen en voor Racismebestrijding v. G4 S Secure Solutions NV*, EU:C:2017:203, European Human Rights Cases, EHRC 2017/96.

[105] Article 16 of the Charter of Fundamental Rights of the European Union.

[106] CJEU, Case C-201/15 AGET Iraklis, ECLI:EU:C:2016:972, par. 86. See also CJEU, Case C-611/12 P, Giordano v Commission, 14 October 2014, EU:C:2014:2282, par. 49.

[107] Joseph Turow, Lauren Feldman, and Kimberly Meltzer, *Open to Exploitation: America's Shoppers Online and Offline*, Annenberg Public Policy Center of the University of Pennsylvania, 2005 http://repository.upenn.edu/cgi/viewcontent.cgi?article=1035&context=asc_papers. ; Joseph Turow et al., *Americans Reject Tailored Advertising and Three Activities that Enable it*,

aims to take consumer protection seriously. For instance, the Charter of Fundamental Rights of the European Union says that 'Union policies shall ensure a high level of consumer protection.'[108] And the Charter grants people a right to non-discrimination.[109] In sum, when balancing the opposing interests (of the store and the consumers), it should be taken into account that the interests on both sides are related to fundamental rights.[110]

Other circumstances can also be considered. For instance, price differences of hundreds of euros could be seen as more problematic than differences of a few euros. And the price differentiation could be seen as more problematic if it concerns necessities such as staple foods, rather than luxury goods. Price differentiation could also be seen as more problematic if a company has a monopoly or similarly powerful position.

To sum up: price differentiation might evade the prohibition of indirect discrimination, if the price differentiation is seen as 'objectively justified by a legitimate aim and the means of achieving that aim are appropriate and necessary'.[111] When price differentiation is objectively justified is hard to say in the abstract. It depends on the circumstances of a case when online price differentiation is prohibited as indirect discrimination. Judges have some leeway to decide either way.

## 4.4     Other protected grounds than ethnic origin

So far, our discussion focused on the Racial Equality Directive. I briefly sketch the relevance of three other non-discrimination directives for online price differentiation. Each of the directives prohibits direct and indirect discrimination.

---

http://ssrn.com/abstract=1478214; Joost Poort and Frederik Zuiderveen Borgesius, *Does everyone have a price? Understanding people's attitude towards online and offline price discrimination*, Internet Policy Review, Volume 8, Issue 1, 2019.
[108] Article 38 of the Charter of Fundamental Rights of the European Union.
[109] Article 21 of the Charter of Fundamental Rights of the European Union.
[110] The phrases "fundamental rights" and "human rights" are used as synonyms in this paper. See on the slight difference: Gloria González Fuster, *The Emergence of Personal Data Protection as a Fundamental Right of the EU* (Dordrecht, Springer 2014), p. 164-166.
[111] Article 2(2)(b) of the Racial Equality Directive 2000/43/EC.

The Gender Goods and Services Directive prohibits discrimination on the basis of gender in the context of the supply of goods and services.[112] Applying this directive (on gender discrimination) would have roughly the same result as applying the Racial Equality Directive. The Gender Goods and Services Directive applies to the 'access to and supply of goods and services',[113] and thus applies to an online store using price differentiation.[114]

However, in the context of stores and price differentiation, EU directives do not protect people against discrimination on the basis of religion or belief, disability, age or sexual orientation. Except for discrimination on the basis of gender and ethnic origin, the EU directives only protect people against discrimination in the employment context. The Employment Equality Directive prohibits discrimination on the basis of religion or belief, disability, age or sexual orientation.[115] This narrow scope of the discrimination directives has been criticised by scholars.[116]

In 2008, the European Commission presented a proposal for a general non-discrimination directive, combatting discrimination on the grounds of religion or belief, disability, age, and sexual orientation.[117] That proposal has a broad scope and applies to the supply of goods and services.[118] Hence, if adopted, the proposal would also apply

---

[112] Gender Goods and Services Directive 2004/113/EC.

[113] Article 1 of the Gender Goods and Services Directive 2004/113/EC.

[114] In other respects, the Gender Goods and Services Directive has a narrower scope than the Racial Equality Directive.

[115] Article 1, Council Directive 2000/78/EC Establishing a General Framework for Equal Treatment in Employment and Occupation, 2000 OJ.L 303/16.

[116] See e.g. Päivi Johanna Neuvonen, *'Inequality in equality' in the European Union equality directives: A friend or a foe of more systematized relationships between the protected grounds?*, International Journal of Discrimination and the Law 2015, Vol. 15(4) 222–240; Lisa Waddington and Mark Bell, *More equal than others: distinguishing European Union equality directives* (2001) 38(3) Common Market Law Review 587; Lisa Waddington, *Future Prospects for EU Equality Law. Lessons to be Learnt from the Proposed Equal Treatment Directive* (2011) 2 European Law Review 163.

[117] Proposal for a Council Directive on implementing the principle of equal treatment between persons irrespective of religion or belief, disability, age or sexual orientation {SEC(2008) 2180} {SEC(2008) 2181} COM/2008/0426 final - CNS 2008/0140. https://eur-lex.europa.eu/legal-content/EN/TXT/HTML/?uri=CELEX:52008PC0426&from=EN

[118] Article 3, idem.

to our price differentiation example. However, after a decade, the proposal is still not adopted.[119]

While the EU non-discrimination directives leave gaps, people in the EU are often protected by law against discrimination on the basis of religion or belief, disability, age or sexual orientation, also outside the employment sector. Many EU member states have laws in place that prohibit discrimination on those grounds.[120]

At first glance, the Charter of Fundamental Rights of the European Union could be read as prohibiting all discrimination, including price differentiation. And recent case law suggests that people can, in certain circumstances, invoke the Charter in horizontal relationships[121]

The Charter states: 'Any discrimination based *on any ground such as* sex, race, colour, ethnic or social origin, genetic features, language, religion or belief, political or any other opinion, membership of a national minority, property, birth, disability, age or sexual orientation shall be prohibited.'[122] The phrase 'any ground such as' shows that the Charter uses an open-ended list of protected grounds.[123]

On reflection, the Charter does not ban *all* discrimination. Such a blanket prohibition of discrimination on the basis of any ground would be absurd. The Charter does – and should – allow a hospital to discriminate on the basis of medical diplomas when hiring a new surgeon. And the Charter allows a country to apply a higher tax rate to people with higher incomes, while such a higher tax rate could be seen, in theory, as discrimination on the basis of income. Hence, the Charter should not be interpreted as prohibiting all types of differentiation and discrimination.

---

[119] See: http://www.europarl.europa.eu/legislative-train/theme-area-of-justice-and-fundamental-rights/file-anti-discrimination-directive
[120] Isabelle Chopin, Carmine Conte and Edith Chambrier, *A comparative analysis of non-discrimination law in Europe 2018. The 28 EU Member States, the former Yugoslav Republic of Macedonia, Iceland, Liechtenstein, Montenegro, Norway, Serbia and Turkey compared,* (2018) European network of legal experts in gender equality and non-discrimination, Publications Office of the European Union, https://www.equalitylaw.eu/downloads/4804-a-comparative-analysis-of-non-discrimination-law-in-europe-2018-pdf-1-02-mb, p. 79-80.
[121] See CJEU (Grand Chamber), Cases C-569/16 and C-570/16, 6 November 2018, ECLI:EU:C:2018:871. See CJEU (Grand Chamber), Cases C-569/16 and C-570/16, 6 November 2018, ECLI:EU:C:2018:871. See also: Chantal Mak, *Case note regarding Bauer and Broßonn, CJEU 6 November 2018, Joined Cases C-569/16 and C-570/16*, European Human Rights Cases 2019/19, p. 57-61 (in Dutch).
[122] Article 21(1) of the Charter of Fundamental Rights of the European Union, emphasis added.
[123] The European Convention on Human Rights also uses an open-ended list of protected characteristics in its non-discrimination provision: Article 14 of the European Convention on Human Rights.

It is not completely clear how an open-ended prohibition of discrimination should be interpreted. But a full discussion of the intricacies of non-discrimination provisions with open-ended lists falls outside the scope of this paper.[124] The next section provides some suggestions for further research.

## 5 Suggestions for further research

In the context of algorithmic decision-making and non-discrimination law, several topics warrant more research and debate. For example, a difficult topic is the conflict between business needs and non-discrimination norms. 'In most jurisdictions,' notes Fredman, 'statutes and case law have specifically permitted individuals or States to defend incursions on equality on the grounds that this is justified as a pursuit of business needs or State macroeconomic policies.'[125] We saw that businesses in the EU can also, under some circumstances, rely on an objective justification, which exempts them from the ban on indirect discrimination. Fredman notes, however, 'that little attention has been paid to why business needs should trump equality.'[126]

What is the right balance between business needs and non-discrimination norms? Legal research can help to answer the question, by identifying and analysing the different arguments. But in the end, it is a largely political question how those arguments are weighed. People who have more trust in the market and capitalism will probably be less inclined to limit business interests and contractual freedom.

The scope of non-discrimination law also calls for more research. Many types of algorithmic differentiation, including differentiation that can be called unfair, fall outside the scope of non-discrimination law. For instance, algorithmic decision-making can reinforce inequality. We saw that online price differentiation can lead to poor people paying higher prices. [127] However, price differentiation (and algorithmic decision-making) could also harm rich people. It is an empirical question whether price differentiation would mostly harm the poor or the rich. Some scholars fear that algorithmic decision-making will mostly harm the poor.[128]

---

[124] See on that topic: Janneke Gerards, *Discrimination grounds*, in Dagmar Schiek, Lisa Waddington and Mark Bell (eds.), *Ius commune case books for a common law of Europe – Non-discrimination* (Oxford: Hart 2007), p. 33-184.
[125] Sandra Fredman, *Discrimination law*, p. 35, (Oxford: Oxford University Press 2011).
[126] Sandra Fredman, *Discrimination law*, p. 35, (Oxford: Oxford University Press 2011).

Most non-discrimination statutes do not protect poor people against such effects. Even if lawmakers wanted to protect poor people against paying more, they would have to tread carefully. Adding 'wealth' as a protected ground to non-discrimination law could have strange effects. If wealth were a protected characteristic, wealthy people could claim that progressive taxing is discriminatory and thus illegal.

The 'tragedy of errors' shows another limitation of non-discrimination law. [129] Algorithmic decision-making can lead to errors in individual cases. Algorithmic decision-making often entails applying a predictive model to individuals. An example of a predictive model is: '90% of the people living in postal code XYZ do not pay attention to prices.' Suppose that an online store, based on this predictive model, raises the prices for consumers in that area. In that case, the company also raises the prices for the 10% who do care about prices. For instance, the postal code could refer to an area with mostly wealthy people, who don't pay much attention to prices. But a poor student renting a room in that street (part of the 10% who does care about prices) will also pay the higher price. Non-discrimination law does not say much about such errors. Whether and how the law should deal with such errors requires more research and debate.

---

[127] See section 2 of this paper.
[128] Virginia Eubanks, *Automating inequality: How high-tech tools profile, police, and punish the poor,* New York: St. Martin's Press, 2018.
[129] The phrase is from: Tal Zarsky, *Mine Your Own Business: Making The Case For The Implications Of The Data Mining Of Personal Information In The Forum Of Public Opinion* (2002) 5 Yale Journal of Law and Technology 1.

## 6    Conclusion

Using online price differentiation as an example, this paper explored to what extent EU non-discrimination law can defend people against unfair algorithmic decision-making. Non-discrimination law, in particular the prohibition of indirect discrimination, can protect people against algorithmic discrimination. Roughly speaking, indirect discrimination occurs when a practice is neutral at first glance but ends up discriminating against people with a protected characteristic, for instance a certain ethnic background. Hence, enforcing non-discrimination law can help to protect people against discriminatory effects of algorithmic decision-making.

But the paper showed that EU non-discrimination law has weaknesses when applied to algorithmic decision-making. First, algorithmic indirect discrimination can remain hidden. Suppose that an online store charges higher prices to people in certain streets, and suppose that this price differentiation practice results in people with a Roma background paying more. Consumers may not realise that they pay more than others. And even if some consumers found out that they pay more than others, they would not yet know that Roma in general pay more, so they would not know about the indirect discrimination. Moreover, it would be difficult for a consumer to prove the indirect discrimination.

Second, the EU non-discrimination directives only protect people against discrimination on the basis of certain protected characteristics (such as ethnicity and gender). However, algorithmic systems can generate *new* categories of people based on seemingly innocuous characteristics, such as web-browser preference, postal code, or more complicated categories combining many data points. An online store may for instance find that most consumers using a certain web browser pay less attention to prices; the store can charge those consumers extra. This type of differentiation could evade non-discrimination law, as browser type is not a protected characteristic, but it could still be unfair.

Third, the prohibition of indirect discrimination results in a nuanced and open norm, which is often difficult to apply in practice. If differentiation is 'objectively justified',[130] the prohibition does not apply. Whether such a justification applies, is context-dependant and requires an intricate proportionality test. Such a nuanced open norm has advantages, but the nuance comes at the cost of clarity.

Fourth, algorithmic decision-making can reinforce social inequality. For example, in some cases, algorithmic pricing has led to higher prices for poor people. The EU non-discrimination directives do not protect people against discrimination on the basis of financial status. Fifth, non-discrimination law is silent about algorithmic decisions based on incorrect predictions, while such decisions can be unfair.

To conclude, current non-discrimination law can be characterised as necessary but not sufficient, as a tool to protect people against unfair algorithmic decisions. Additional regulation is probably necessary.[131] More research and debate are needed on the question of how people should be protected against algorithmic discrimination.

* * *

---

[130] See for instance article 2(2)(b) of the Racial Equality Directive (2000/43/EC).
[131] See for a possible way to develop rules for algorithmic decision-making: chapter VI.3 of Frederik Zuiderveen Borgesius, *Discrimination, artificial intelligence, and algorithmic decision-making*, report for the European Commission against Racism and Intolerance (ECRI), Council of Europe, 2019, https://www.coe.int/en/web/artificial-intelligence/-/news-of-the-european-commission-against-racism-and-intolerance-ecri-